\documentclass[11pt,times]{emulateapj}

\usepackage{txfonts}
\usepackage{graphicx}
\usepackage{amssymb}
\usepackage{epsfig}
\usepackage{natbib}
\usepackage{textcomp}
\usepackage{color}

\newcommand{\cold}[2]{$#1\times10^{#2}$ cm$^{-2}$}

\newcommand{\be}{\begin{equation}}
\newcommand{\ee}{\end{equation}}

\newcommand{\eg}{\emph{e.g.}}
\newcommand{\kms}{\mbox{km\,\ensuremath{\rm{s}^{-1}}}}

\submitted{ApJ, in press}

\shorttitle{7~mm spectroscopy of the young protostar Chamaeleon MMS1}
\shortauthors{Cordiner et al.}

\begin{document}

\title{Organic chemistry of low-mass star-forming cores I:\\ 7~mm spectroscopy of Chamaeleon MMS1}

\author{Martin A. Cordiner\altaffilmark{1}, Steven B. Charnley, Eva S. Wirstr{\"o}m\altaffilmark{1}}
\affil{Astrochemistry Laboratory and The Goddard Center for Astrobiology, Mailstop 691, NASA Goddard Space Flight Center, 8800 Greenbelt Road, Greenbelt, MD 20770, USA}
\email{martin.cordiner@nasa.gov}
\and
\author{Robert  G. Smith}
\affil{School of Physical, Environmental \& Mathematical Sciences, The University of New South Wales, Australian Defence Force Academy, Canberra ACT 2600, Australia}

\altaffiltext{1}{Department of Physics, The Catholic University of America, Washington, DC 20064, USA}

\keywords{Astrochemistry -- Stars: formation -- ISM: abundances -- Radio lines: ISM, stars}

\begin{abstract}

Observations are presented of emission lines from organic molecules at frequencies $32-50$~GHz in the vicinity of Chamaeleon MMS1. This chemically-rich dense cloud core habours an extremely young, very low-luminosity protostellar object and is a candidate first hydrostatic core. Column densities are derived and emission maps are presented for species including polyynes, cyanopolyynes, sulphuretted carbon-chains and methanol. The polyyne emission peak lies about 5000~AU from the protostar, whereas methanol peaks about 15,000~AU away. Averaged over the telescope beam, the molecular hydrogen number density is calculated to be $10^6$~cm$^{-3}$ and the gas kinetic temperature is in the range $5-7$~K.  The abundances of long carbon chains are very large, and are indicative of a non-equilibrium carbon chemistry; C$_6$H and HC$_7$N column densities are \cold{5.9^{+2.9}_{-1.3}}{11} and \cold{3.3^{+8.0}_{-1.5}}{12}, respectively, which are similar to the values found in the most carbon-chain-rich protostars and prestellar cores known, and are unusually large for star-forming gas. Column density upper limits were obtained for the carbon-chain anions C$_4$H$^-$ and C$_6$H$^-$, with anion-to-neutral ratios [C$_4$H$^-$]/[C$_4$H]~$<$~0.02\% and [C$_6$H$^-$]/[C$_6$H]~$<$~10\%, consistent with previous observations in interstellar clouds and low-mass protostars.  Deuterated HC$_3$N and $c$-C$_3$H$_2$ were detected. The [DC$_3$N]/[HC$_3$N] ratio of approximately 4\% is consistent with the value typically found in cold interstellar gas.

\end{abstract}

\maketitle

\section{Introduction}
\label{sec:int}

During the earliest stages of star formation, quiescent molecular gas collapses to form a dense, prestellar core and subsequently a protostar with accretion disk, from which planetary systems may eventually arise. Understanding the evolution of molecular complexity as matter passes through these phases is crucial to the development of theories concerning organic chemistry during the formation of planetary systems, with relevance to the origin of life in the Galaxy \citep{ehr00}. Interstellar clouds and prestellar cores have been shown to host a rich organic chemistry that results in the synthesis of a range of large organic molecules \citep{her09}. Unsaturated carbon-chain-bearing species such as the polyynes and cyanopolyynes are highly abundant in dark interstellar clouds such as TMC-1 \citep[\eg][]{cer84}, and have recently been found to be present in large concentrations around some prestellar cores and low-mass protostars \citep{sak08,sak09,gup09,cor11}. It has been hypothesised that such complex molecules may survive the passage from protostellar envelope to accretion disk, to later become incorporated, largely unmodified, into solar system bodies such as comets \citep[\eg][]{mei98,wil07,cha08,char09,mum11}.  Further observational and theoretical studies of the chemical evolution of gas and dust in the vicinity of protostars as they form and heat their surroundings will be crucial in order to understand the chemical reagents present at the start of planet formation, and those delivered to planetary surfaces later by cometary impacts.

Chamaeleon MMS1 (abbreviated Cha-MMS1) is an ideal source in which to study the chemistry of a protostar in its natal environment. Thus far, Cha-MMS1 has been relatively little-studied in terms of its molecular inventory. It was first detected in 1.3~mm emission as a very compact object ($\approx 30''$ in diameter) by \citet{rei96}. Infrared and sub-mm photometry are consistent with the presence of a young Class 0 protostar embedded in a dusty envelope with a core density of 3.3$\times10^6$~cm$^{-3}$ and diameter of about 3600~AU \citep{bel06}. The central object has been classified by \citet{bel06,bel11} as either an extremely young, very low-luminosity Class~0 protostar (VeLLO) or an example of a first hydrostatic core (FHSC). VeLLOs are likely to be young protostars undergoing episodic accretion, and a number of these were identified by the recent Spitzer c2d survey \citep{dun08}.  The FHSC is a transient, quasi-hydrostatic object predicted by the simulations of \citet{lar69}, but so far eluding observational confirmation. It is theorised to be formed in the early stages of protostellar collapse at the point when the density becomes sufficiently high that the core is opaque to thermal radiation. Virial heating then provides the energy to pressurise the (predominantly molecular hydrogen) gas against further collapse. Representative of a phase of protostar evolution preceding Class~0, the FHSC is also known as Class~$-$I \citep{bos95}. The physics and chemistry of FHSCs and VeLLOs have recently been discussed by \citet{omu07} and \citet{lee07}, respectively.  The predicted spectral energy distribution of an FHSC is similar to that of the coolest, lowest-luminosity VeLLOs, and at present there are only four likely candidate FHSCs reported in the literature: Cha-MMS1, L1448~IRS2E \citep{che10}, Per-Bolo~58 \citep{eno10} and L1451-mm \citep{pin11}.

Observations of organic molecules in Cha-MMS1 were made previously by \citet{kon00} using SEST; \citet{bel06} observed a selection of deuterated and undeuterated molecular ions, and \citet{ten06} mapped the HCN and HNC emission around the protostar. However, the chemical and physical characterisation of Cha-MMS1 is still far from complete. As part of a program to understand organic chemistry during the earliest phases of low-mass star formation, we have initiated molecular line surveys in a sample of around 70 young protostars, focussing on individual sources for more detailed followup. In this article, we present 7~mm spectral line data and maps of cold carbon-chain-bearing species and other organic molecules in the vicinity of Cha-MMS1. These observations are analysed in the context of the physical and chemical properties of the protostellar envelope which, compared with other young  carbon-chain-rich protostars (such as L1527 and L1521F, whose organic chemistry were recently examined by \citealt{sak08} and \citealt{tak11}), help to constrain the current state and the history of the core.

\section{Observations}

\label{sec:obs}

\begin{figure}
\centering
\includegraphics[width=0.48\columnwidth]{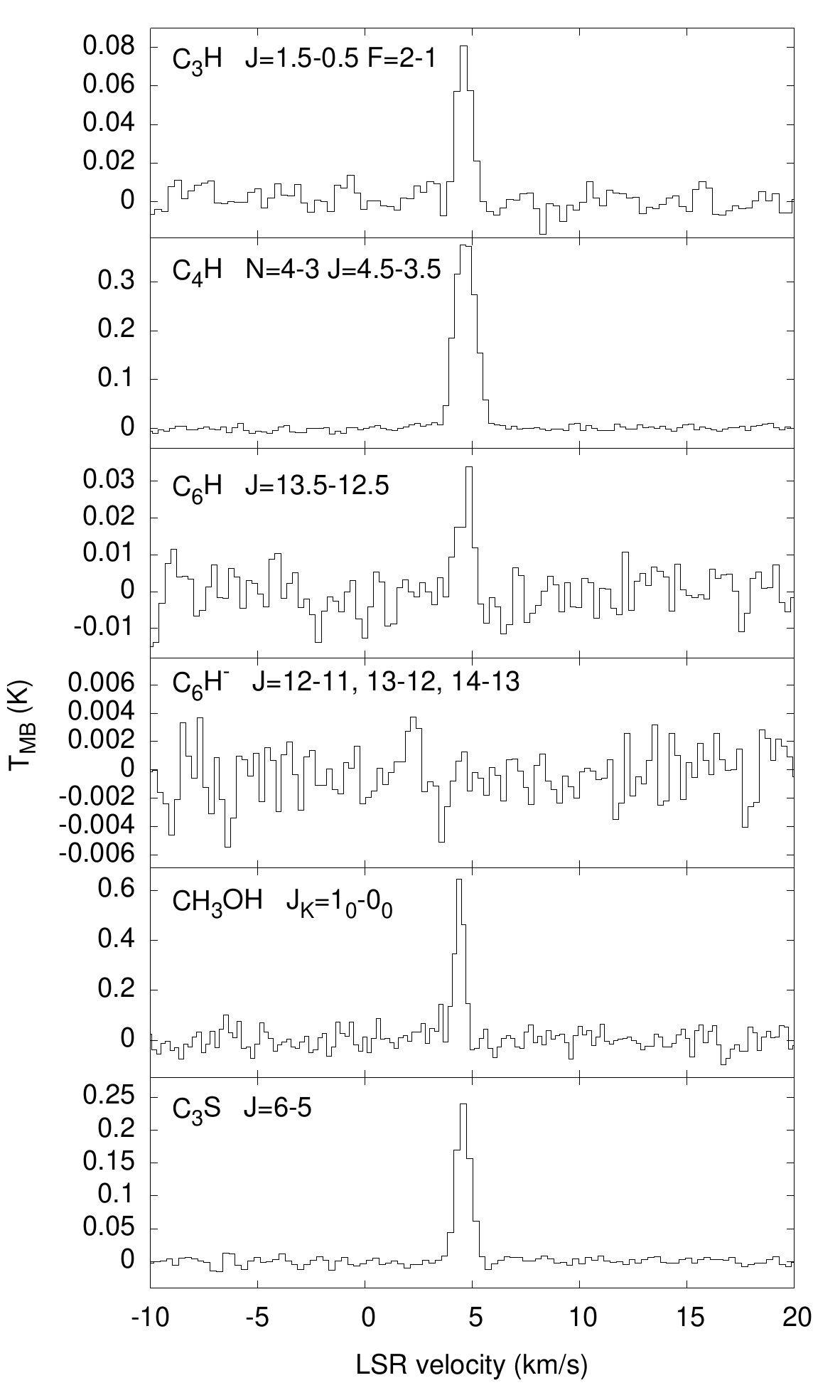}
\includegraphics[width=0.48\columnwidth]{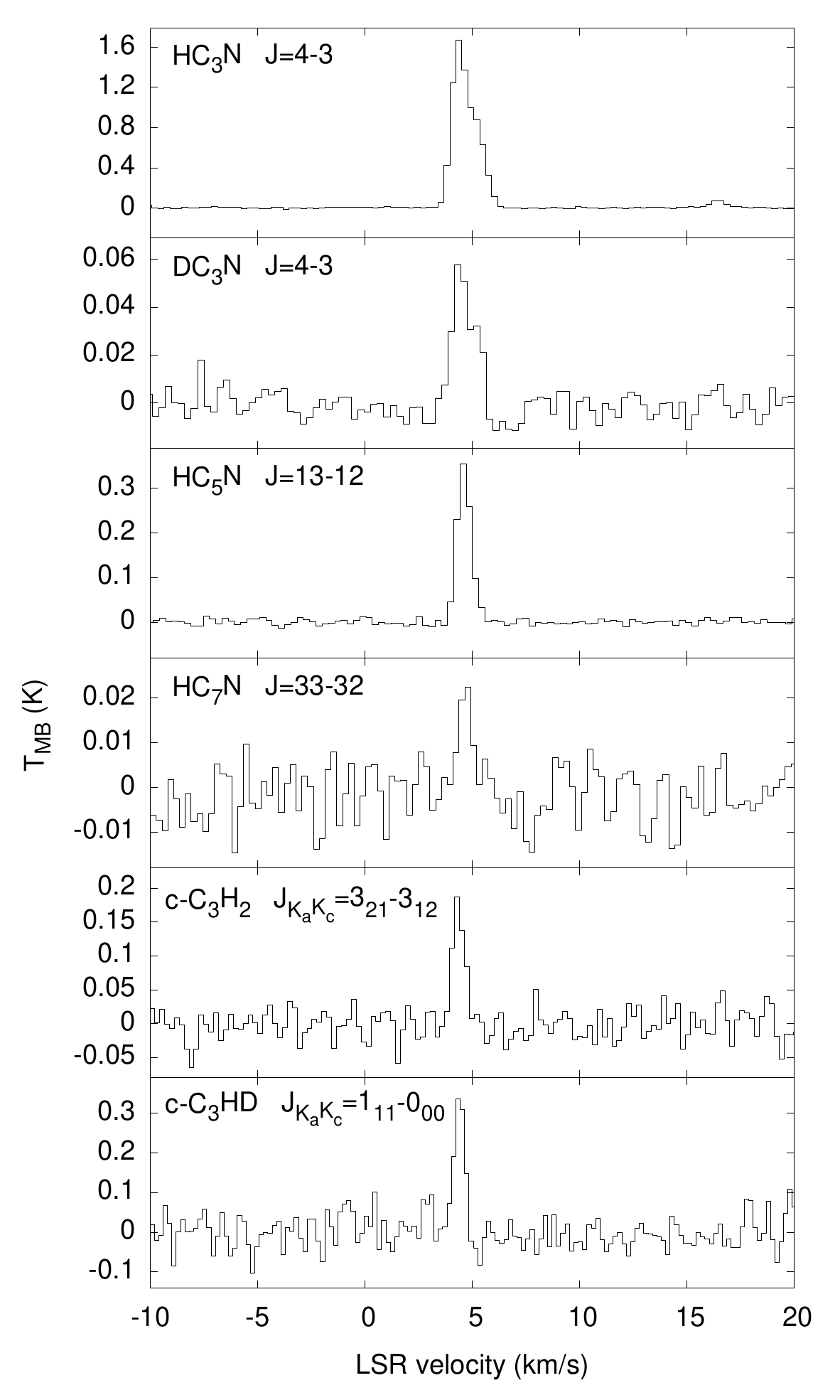}
\caption{Example 7~mm Mopra spectra of Cha-MMS1. For C$_6$H$^-$, the spectrum shown is the average of the three observed lines.}
\label{fig:spectra}
\end{figure}

Observations of Cha-MMS1 were carried out in March 2010 using the ATNF Mopra 22~m radio telescope. The Mopra Spectrometer (MOPS) was configured in zoom mode to simultaneously observe 16 spectral windows distributed over an 8~GHz range, each with a bandwidth of 137.5~MHz and a channel spacing of 33.6~kHz. During observations, the total system temperatures were in the range $80-200$~K. The beam efficiency factor at 7~mm (42~GHz) was 0.6. Pointing was checked about every three hours and was typically accurate to within $5-10''$. 

Position-switched integrations of Cha-MMS1 were performed at the \citep{kon00} peak HC$_3$N position, at (J2000) coordinates R.A. = 11:06:33, decl. = $-77$:23:46.  The off position used was R.A. = 11:17:38, decl. = $-77$:23:25. To help mitigate the known Mopra $30$~MHz spectral baseline ripple, the telescope subreflector was continuously driven up and down during observations, with a displacement of 3.5~mm. Atmospheric opacity corrections $e^{\tau_{\theta}}$ are required at 7~mm; zenith optical depths as a function of frequency ($\tau_0(\nu)$) were calculated using the GILDAS ATM routine\footnote{http://www.iram.fr/IRAMFR/GILDAS} based on the precipitable water vapour column and atmospheric temperature and pressure at the time of observations.

Observations were performed using two different MOPS configurations centered at frequencies of 36~GHz and 45~GHz, respectively.  Approximately 22~hr of on-source integration was obtained for the 36~GHz setting, yielding RMS main-beam brightness temperature sensitivities $T_{\rm MB}({\rm RMS})\approx6$~mK per channel. Two hours integration in the 45~GHz setting yielded $T_{\rm MB}({\rm RMS})\approx30$~mK per channel. The observed molecular transition lines are given in Table \ref{tab:lines}. The telescope beam FWHM was $96''$ at 36~GHz and $77''$ at 45~GHz.

\begin{deluxetable*}{lllllll}
\tabletypesize{\scriptsize}
\tablecaption{Observed line parameters\label{tab:lines}}
\tablewidth{0pt}
\tablehead{
\colhead{Species}& \colhead{Transition} & \colhead{Freq.} & \colhead{$T_{\rm MB}$} & \colhead{$v_{\rm LSR}$} & \colhead{$\Delta v$} & \colhead{$\int T_{\rm MB}\,dv$}\\
&&\colhead{(MHz)} & \colhead{(mK)} & \colhead{(\kms)} & \colhead{(\kms)} & \colhead{(mK\,\kms)}
}
\startdata

C$_6$H&$^2\Pi_{3/2}\ J=23/2-21/2\ f$&31881.8596&17 (7)&4.71 (15)&0.93 (34)&17 (6)\\
C$_6$H&$^2\Pi_{3/2}\ J=23/2-21/2\ e$&31885.5414&21 (6)&4.58 (13)&1.08 (31)&25 (7)\\
C$_3$H&$J=3/2-1/2\ e\ F=2-1$&32660.6550&86 (10)&4.75 (2)&0.78 (6)&72 (8)\\
C$_3$H&$J=3/2-1/2\ e\ F=1-0$&32663.3750&39 (7)&4.86 (5)&0.79 (14)&32 (6)\\
C$_3$H&$J=3/2-1/2\ e\ F=1-1$&32667.6366&22 (7)&4.37 (12)&0.86 (27)&20 (6)\\
HC$_7$N&$J=29-28$&32711.6820&19 (11)&4.70 (9)&0.47 (24)&14 (6)\\
C$_6$H$^-$&$J=12-11$&33044.4879&$<$16& \nodata& \nodata&$<$13\\
CC$^{34}$S&$N_J=2_3-1_2$&33111.8390&34 (7)&4.76 (7)&0.94 (16)&37 (7)\\
CCS&$N_J=2_3-1_2$&33751.3737&792 (45)&4.81 (1)&0.76 (1)&650 (38)\\
DC$_3$N&$J=4-3$&33772.5312&58 (7)&4.71 (5)&1.3 (10)&80 (9)\\
HC$_7$N&$J=30-29$&33839.6250&$<$19& \nodata& \nodata&$<$16\\
HC$_5$N&$J=13-12$&34614.3870&368 (22)&4.76 (1)&0.77 (1)&301 (19)\\
C$_6$H&$^2\Pi_{3/2}\ J=25/2-23/2\ f$&34654.0373&13 (6)&4.68 (22)&1.28 (48)&22 (6)\\
C$_6$H&$^2\Pi_{3/2}\ J=25/2-23/2\ e$&34658.3831&14 (5)&4.71 (20)&1.39 (49)&18 (5)\\
C$_3$S&$J=6-5$&34684.3675&247 (17)&4.70 (1)&0.76 (2)&200 (15)\\
C$_6$H&$^2\Pi_{1/2}\ J=27/2-25/2\ f$&34887.1145&$<$18& \nodata& \nodata&$<$15\\
C$_6$H&$^2\Pi_{1/2}\ J=27/2-25/2\ e$&34917.8847&$<$18& \nodata& \nodata&$<$15\\
HC$_7$N&$J=31-30$&34967.5880&11 (6)&4.75 (27)&1.52 (65)&17 (5)\\
C$_6$H$^-$&$J=13-12$&35798.1532&$<$17 \nodata& \nodata& \nodata&$<$14 \nodata\\
C$_4$H$_2$&$J_{K_aK_c}=4_{13}-3_{12}$ (ortho)&35875.7746&33 (6)&4.65 (7)&1.14 (18)&41 (7)\\
HC$_7$N&$J=32-31$&36095.5340&24 (12)&4.67 (7)&0.39 (17)&10 (4)\\
CH$_3$OH&$J_k=4_{-1}-3_0\ {\rm E}$&36169.2610&94 (10)&4.71 (2)&0.87 (5)&86 (9)\\
HC$_3$N&$J=4-3\ F=4-4$&36390.8861&76 (9)&4.77 (3)&0.82 (6)&83 (9)\\
HC$_3$N\tablenotemark{a}&$J=4-3\ \Delta F=-1$&36392.3240&1673 (90)& \nodata& \nodata&2158 (114)\\
HC$_3$N&$J=4-3\ F=3-43$&36394.1765&83 (9)&4.74 (2)&0.74 (5)&65 (7)\\
HC$_7$N&$J=33-32$&37223.4920&24 (7)&4.82 (9)&0.67 (20)&19 (6)\\
C$_4$H$^-$&$J=4-3$&37239.4020&$<$17& \nodata& \nodata&$<$14\\
HC$_5$N&$J=14-13$&37276.9940&264 (18)&4.77 (1)&0.79 (2)&223 (16)\\
C$_6$H&$^2\Pi_{3/2}\ J=27/2-25/2\ f$&37426.1916&31 (8)&4.87 (6)&0.69 (15)&24 (5)\\
C$_6$H&$^2\Pi_{3/2}\ J=27/2-25/2\ e$&37431.2550&19 (5)&4.61 (14)&1.5 (33)&32 (7)\\
C$_4$H&$N=4-3\ J=9/2-7/2$&38049.6583&398 (24)&4.80 (1)&1.15 (1)&494 (32)\\
C$_4$H&$N=4-3\ J=7/2-5/2$&38088.4587&385 (23)&4.78 (1)&0.93 (1)&384 (24)\\
CH$_3$CHO&$J_{K_aK_c}=2_{02}-2_{01}\ {\rm E}$&38505.9990&44 (6)&4.50 (5)&0.87 (11)&40 (6)\\
CH$_3$CHO&$J_{K_aK_c}=2_{02}-2_{01}\ {\rm A}$&38512.1130&59 (8)&4.97 (3)&0.83 (8)&55 (7)\\
C$_6$H$^-$&$J=14-13$&38551.8083&$<$17& \nodata& \nodata&$<$14\\
DC$_3$N&$J=5-4$&42215.5827&108 (17)&4.59 (5)&0.95 (12)&110 (18)\\
HCO$_2^+$&$J_{K_aK_c}=2_{02}-1_{01}$&42766.1975&$<$51& \nodata& \nodata&$<$42\\
$t$-CH$_3$CH$_2$OH&$J_{K_aK_c}=1_{01}-0_{00}$&43026.6000&$<$51& \nodata& \nodata&$<$42\\
DC$_5$N&$J=17-16$&43215.4382&$<$51& \nodata& \nodata&$<$42\\
HNCO&$J_{K_aK_c}=2_{02}-1_{01}$&43963.0000&100 (24)&4.45 (7)&0.72 (17)&81 (24)\\
CCS&$N_J=4_3-3_2$&43981.0270&127 (30)&4.70 (5)&0.54 (11)&86 (20)\\
$c$-C$_3$H$_2$&$J_{K_aK_c}=3_{21}-3_{12}$ (ortho)&44104.7806&189 (30)&4.48 (3)&0.63 (8)&128 (21)\\
C$_4$D&$N=5-4\ J=11/2-9/2$&44146.0772&$<$63& \nodata& \nodata&$<$52\\
C$_4$D&$N=5-4\ J=9/2-7/2$&44182.0627&$<$63& \nodata& \nodata&$<$52\\
HC$_5$N&$J=17-16$&45264.7199&278 (34)&4.53 (2)&0.62 (5)&193 (25)\\
HC$_3$N\tablenotemark{a}&$J=5-4\ \Delta F=-1$&45490.3138&2198 (132)& \nodata& \nodata&1965 (115)\\
$c$-C$_3$H$_2$&$J_{K_aK_c}=2_{11}-2_{02}$ (para)&46755.6136&840 (63)&4.56 (1)&0.64 (2)&618 (54)\\
$t$-CH$_3$CH$_2$OH&$J_{K_aK_c}=4_{04}-3_{13}$&46832.8000&$<$81& \nodata& \nodata&$<$67\\
C$_4$H&$N=5-4\ J=11/2-9/2$&47566.8090&635 (54)&4.68 (1)&0.64 (3)&443 (38)\\
C$_4$H&$N=5-4\ J=9/2-7/2$&47605.4850&515 (46)&4.55 (1)&0.65 (3)&380 (39)\\
HC$_5$N&$J=18-17$&47927.2746&226 (43)&4.52 (3)&0.45 (7)&106 (21)\\
CH$_3$OH&$J_k=1_0-0_0\ {\rm A}^+$&48372.4600&657 (68)&4.51 (1)&0.52 (3)&365 (37)\\
CS&$J=1-0$&48990.9549&2465 (145)&4.44 (1)&0.91 (1)&2579 (160)\\
C$_3$N\tablenotemark{a}&$N=5-4\ J=11/2-9/2$&49466.4200&117 (42)&4.72 (9)&0.68 (24)&77 (21)\\
C$_3$N&$N=5-4\ J=9/2-7/2$&49485.2480&$<$126& \nodata& \nodata&$<$104\\
$c$-C$_3$HD&$J_{K_aK_c}=1_{11}-0_{00}$&49615.8517&366 (54)&4.48 (3)&0.56 (6)&222 (34)\\
\enddata
\tablenotetext{a} {HC$_3$N and C$_3$N measurements are integrated over the three main hyperfine components.}
\tablecomments{Errors on the last quoted digit(s) are given in parentheses. Upper limits are $3\Delta v\times{\rm RMS\ noise}$, with $\Delta v=0.78$.}
\end{deluxetable*}

On-the-fly mapping of the $8'\times8'$ region surrounding Cha-MMS1 was performed using the 36~GHz setting, employing a Nyquist sampling rate in the R.A. (scan) direction and $33''$ spacing between map rows. The resulting map sensitivity was $T_{\rm MB}({\rm RMS})\approx70$~mK per channel per pixel.

Spectral line data were reduced and combined using standard routines in the ATNF Spectral Analysis Package (ASAP) version 3.0. Baseline subtraction and spectral line analysis were performed using NRAO IRAF\footnote{IRAF is distributed by the National Optical Astronomy Observatory, which is operated by the Association of Universities for Research in Astronomy (AURA) under cooperative agreement with the National Science Foundation.}. Examples of a selection of the observed spectral lines are shown in Figure \ref{fig:spectra}.  Map data were reduced using the AIPS++ programs LIVEDATA and GRIDZILLA \citep[see][]{bar01}, and smoothed with a Gaussian FWHM of $96''$ applicable to the beam size at 36~GHz. 

\section{Results}
\label{sec:res}

\subsection{Molecular emission maps}

\begin{figure*}
\centering
\includegraphics[width=0.9\columnwidth]{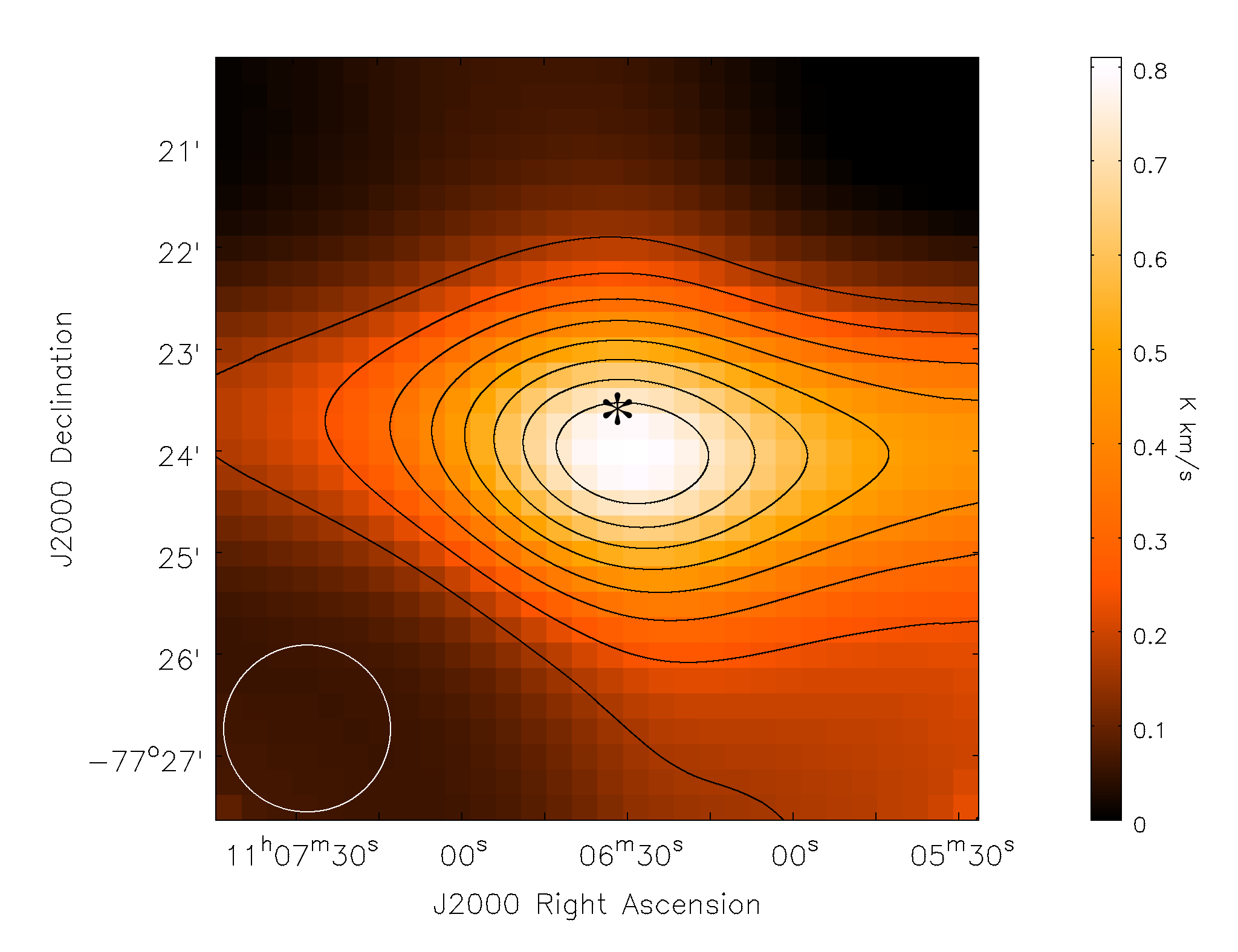}
\includegraphics[width=0.9\columnwidth]{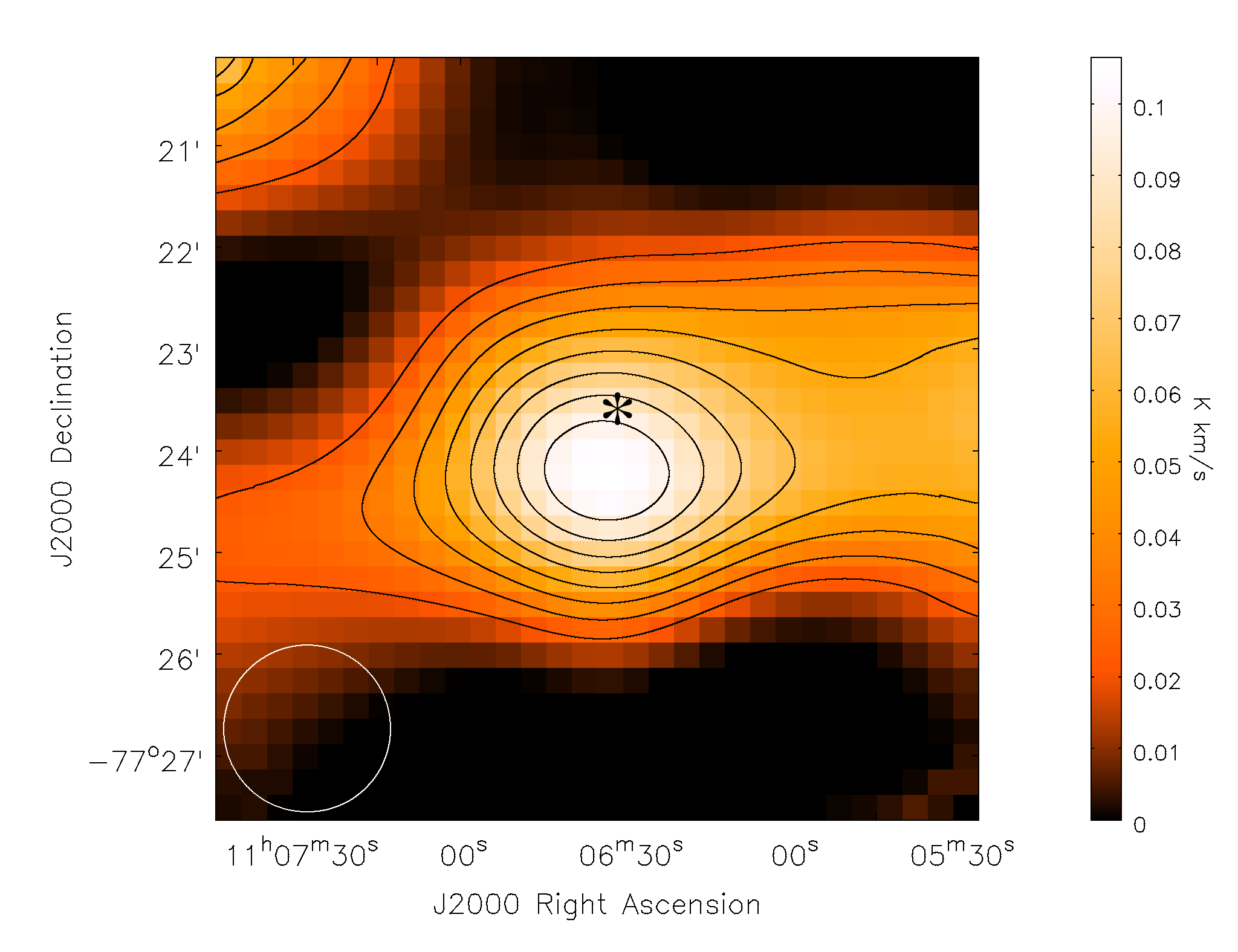}
\caption{Integrated emission intensity maps of HC$_3$N $J=4-3$ (left) and HC$_5$N $J=13-12$ (right). The protostar Cha-MMS1 is marked with an asterisk, which denotes the center of the millimeter source. Telescope beam shown in bottom left.}
\label{fig:hc3n_map}
\end{figure*}

\begin{figure*}
\centering
\includegraphics[width=0.9\columnwidth]{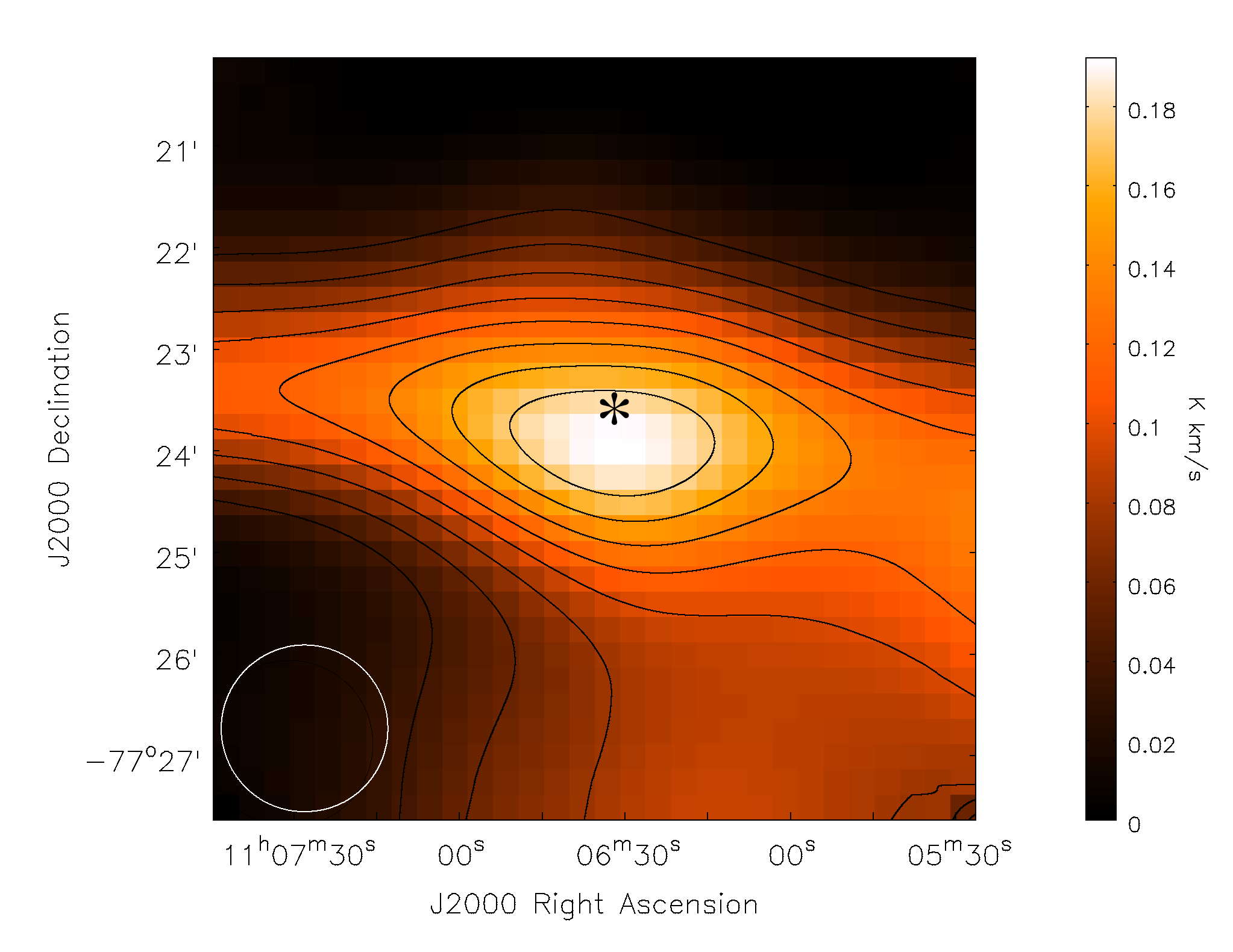}
\includegraphics[width=0.9\columnwidth]{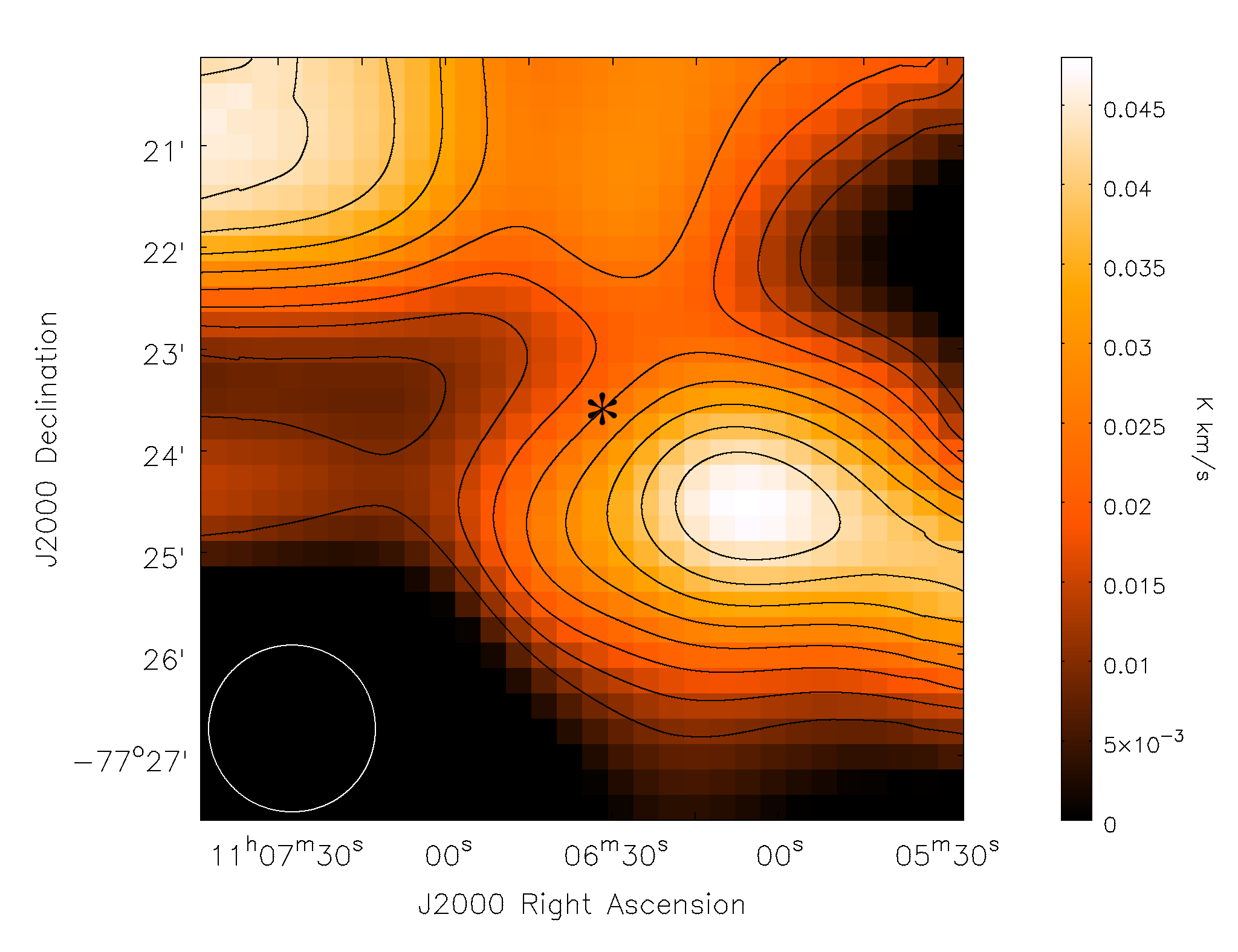}
\caption{Integrated emission intensity maps of C$_4$H $N=4-3$ (left) and CH$_3$OH $J_k=4_{-1}-3_0 {\rm E}$ (right). The protostar Cha-MMS1 is marked with an asterisk, which denotes the center of the millimeter source. Telescope beam shown in bottom left.}
\label{fig:ch3oh_map}
\end{figure*}

\begin{figure*}
\centering
\includegraphics[width=0.9\columnwidth]{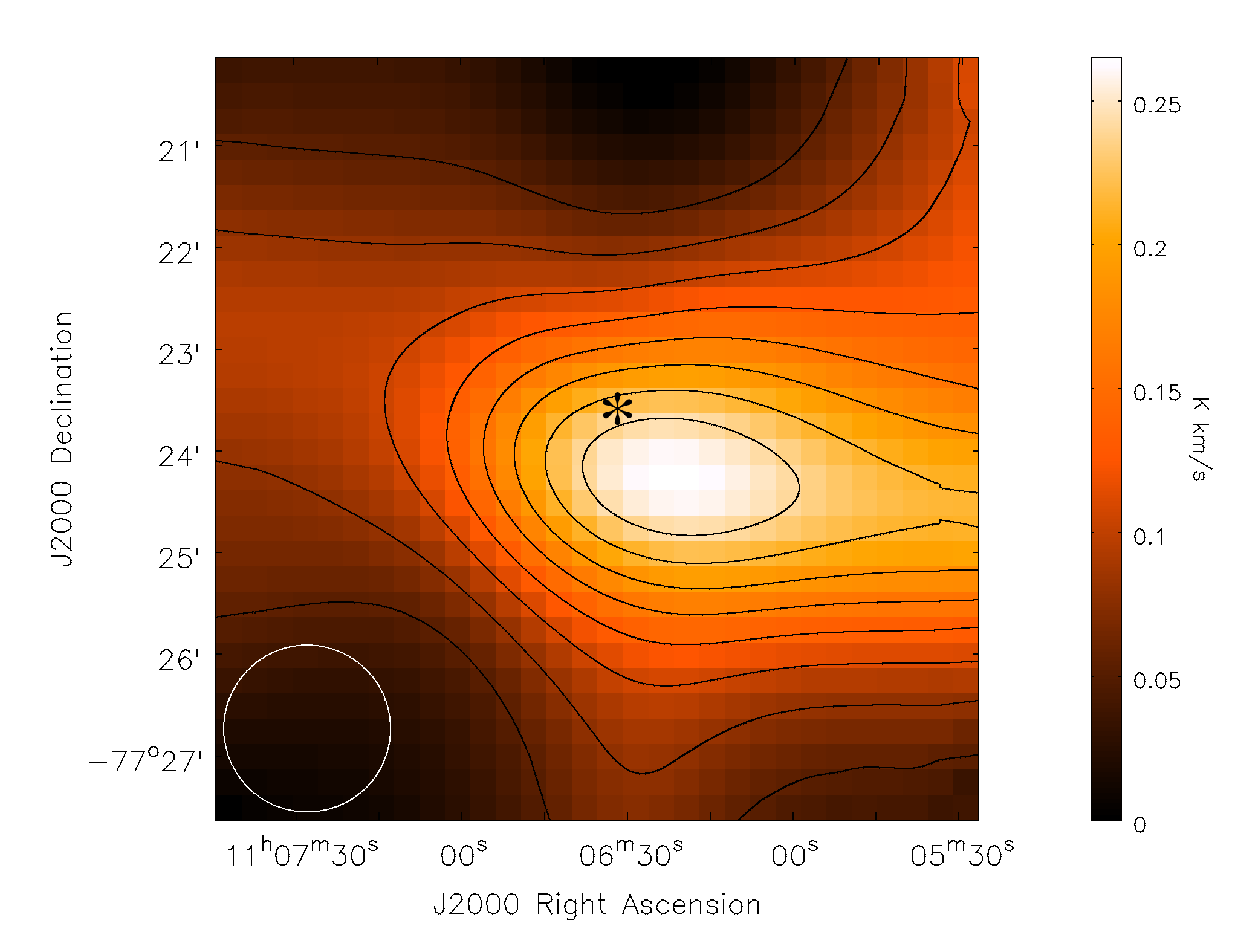}
\includegraphics[width=0.9\columnwidth]{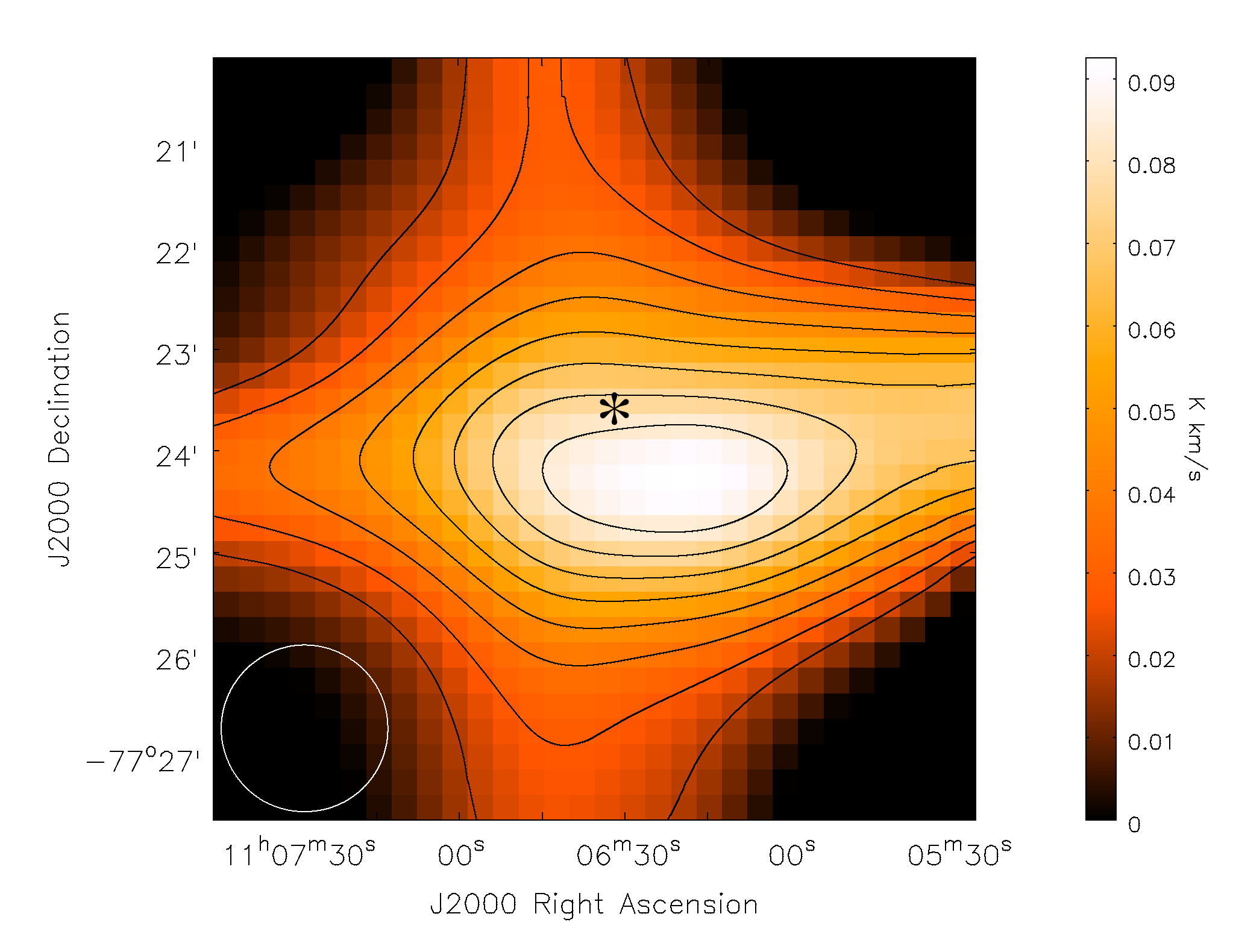}
\caption{Integrated emission intensity maps of CCS $N_J=4_3-3_2$ (left) and C$_3$S $J=6-5$ (right). The protostar Cha-MMS1 is marked with an asterisk, which denotes the center of the millimeter source. Telescope beam shown in bottom left.}
\label{fig:c3s_map}
\end{figure*}

Maps of the integrated line intensity $\int T_{\rm A}\,dv$ are shown for HC$_3$N, HC$_5$N, C$_4$H, CH$_3$OH, CCS and C$_3$S in Figures \ref{fig:hc3n_map} to \ref{fig:c3s_map}. All mapped species show resolved structure with strong emission peaks within $2'$ of the protostar (which is plotted as an asterisk).  The polyyne-based species HC$_3$N, HC$_5$N and C$_4$H all show peaks that are approximately spatially coincident with each other and are elongated in the east-west direction. The sulphuretted carbon chains show a similar morphology, but are offset by about $35''$ to the west of the polyynes.  Elliptical Gaussian fits to these carbon-chain-bearing species have ${\rm FWHM}\sim200''\times400''$, with the long axis orientated at a mean position angle $-5^\circ$ from the horizontal (E-W) axis. The methanol peak has similar dimensions, but is located $70''$ south-west of the polyyne peak. Extended structures are present in all the carbon chain maps, the principal common feature being a ridge in the east-west direction, the full extent of which cannot be determined from our maps.  The methanol distribution is markedly different however, with relatively little extension to the east of the protostar, and a secondary peak to the north-east.  The polyyne-based species peak closest to the protostar (about $30-40''$ away), followed by the sulphuretted carbon chains ($60''$ away), then methanol ($100''$ away).  Assuming a distance to Cha-MMS1 of 150~pc \citep{knu98}, these correspond to respective distances from the protostar of about 5,000~AU, 9,000~AU and 15,000 AU in the plane of the sky, and indicate a gradient in the chemical/physical properties of the gas.

\subsection{Molecular column densities}

All spectral lines -- observed at the \citet{kon00} HC$_3$N peak -- were fitted using single Gaussians. Integrated line intensities and fit parameters are given in Table \ref{tab:lines}.  One-sigma errors were calculated using Monte Carlo noise re-sampling with 1000 replications.  The errors in $T_{\rm MB}$ and $\int T_{\rm MB}\,dv$ given in the table also include uncertainties of $\pm5$\% to account for errors in $T_{\rm MB}$ introduced by the atmospheric opacity correction as a result of uncertainties in the atmospheric model parameters. Measured line FWHM are in the range $0.4-1.5$ \kms.  Unresolved hyperfine structure is present in many of the lines, and causes significant additional broadening of the observed line widths for HC$_3$N, C$_3$N, DC$_3$N, C$_4$H and C$_6$H.  Ignoring outliers more than one standard deviation from the mean, the average line FWHM is 0.78 \kms.

Molecular column densities are presented in Table \ref{tab:colds}.  These were calculated using a variety of methods as indicated in the table notes. Spectral line data and partition functions have been taken from The Cologne Database for Molecular Spectroscopy \citep{mul01}, with supplementary data from the JPL database\footnote{http://spec.jpl.nasa.gov/}.  Partition functions for the different nuclear spin configurations of $c$-C$_3$H$_2$ and CH$_3$OH were calculated by direct summation over the relevant energy levels. The beam-filling factor was assumed to be unity for all emission, which is justified based on the spatial extent of the emission maps. However, structure on size scales less than the size of the Mopra beam is likely given the presence of the compact protostellar core. This may introduce additional uncertainties into the derived column densities and temperatures.

\begin{deluxetable}{llllc}
\tabletypesize{\small}
\tablecaption{Molecular column densities and abundances\label{tab:colds}}
\tablewidth{0pt}
\tablehead{
\colhead{Species} & \colhead{$N$ (cm$^{-2}$)}& \colhead{$T$ (K)} & \colhead{Abundance} & \colhead{Method\tablenotemark{a}}
}
\startdata
C$_3$H&$2.9^{+0.9}_{-0.5}\,(12)$&$6.1\pm1.4$&$9.5^{+10}_{-3.6}\,(-11)$&(1)\\[2pt]
C$_4$H&$1.4^{+0.8}_{-0.5}\,(14)$&$4.9\pm1.0$&$4.6^{+6.6}_{-2.3}\,(-9)$&(2)\\[2pt]
C$_4$H$^-$&$<4.5\,(10)$&$6.1\pm1.4$&$<2.4\,(-12)$&(1)\\[2pt]
C$_6$H&$5.9^{+2.9}_{-1.3}\,(11)$&$6.1\pm1.4$&$2.0^{+2.7}_{-0.9}\,(-11)$&(1)\\[2pt]
C$_6$H$^-$&$<4.8\,(10)$&$6.1\pm1.4$&$<2.5\,(-12)$&(1)\\[2pt]
C$_3$N&$2.5^{+1.6}_{-0.9}\,(12)$&$6.1\pm1.4$&$8.2^{+13}_{-4.3}\,(-11)$&(1)\\[2pt]
HC$_3$N&$2.0\,(13)$&$7.1$&$6.7\,(-10)$&(3)\\[2pt]
HC$_3$N&$1.6^{+0.2}_{-0.1}\,(13)$&$7.2\pm0.2$&$5.3^{+3.6}_{-3.8}\,(-10)$&(4)\\[2pt]
HC$_5$N&$4.5^{+1.6}_{-1.2}\,(12)$&$7.1\pm0.8$&$1.5^{+1.7}_{-0.7}\,(-10)$&(2)\\[2pt]
HC$_7$N&$3.3^{+8.0}_{-1.5}\,(12)$&$6.1\pm1.4$&$1.1^{+4.9}_{-0.7}\,(-10)$&(1)\\[2pt]
DC$_3$N&$7.3^{+2.3}_{-1.1}\,(11)$&$6.1\pm1.4$&$2.4^{+2.6}_{-0.9}\,(-11)$&(2)\\[2pt]
DC$_5$N&$<4.9\,(12)$&$6.1\pm1.4$&$<2.6\,(-10)$&(1)\\[2pt]
o-$c$-C$_3$H$_2$&$1.3^{+5.9}_{-0.4}\,(13)$&$6.1\pm1.4$&$4.3^{+31}_{-2.0}\,(-10)$&(3)\\[2pt]
o-$c$-C$_3$H$_2$&$8.7^{+0.9}_{-0.8}\,(12)$&$6.6\pm0.3$&$2.9^{+2.1}_{-0.9}\,(-10)$&(4)\\[2pt]
p-$c$-C$_3$H$_2$&$1.4^{+1.0}_{-0.3}\,(13)$&$6.1\pm1.4$&$4.7^{+7.3}_{-2.0}\,(-10)$&(3)\\[2pt]
p-$c$-C$_3$H$_2$&$1.5^{+1.1}_{-0.4}\,(13)$&$6.1\pm1.4$&$5.0^{+5.5}_{-1.0}\,(-10)$&(1)\\[2pt]
$c$-C$_3$HD&$4.5^{+1.3}_{-0.7}\,(12)$&$6.1\pm1.4$&$1.5^{+1.5}_{-0.6}\,(-10)$&(1)\\[2pt]
C$_4$H$_2$\tablenotemark{b}&$2.1^{+3.5}_{-1.3}\,(12)$&$6.1\pm1.4$&$6.8^{+23}_{-4.9}\,(-11)$&(1)\\[2pt]
HNCO&$7.6^{+3.5}_{-2.3}\,(12)$&$6.1\pm1.4$&$2.5^{+3.3}_{-1.2}\,(-10)$&(1)\\[2pt]
HCO$_2^+$&$<9.0\,(11)$&$6.1\pm1.4$&$<4.7\,(-11)$&(1)\\[2pt]
CH$_3$OH-A&$6.0^{+1.9}_{-1.3}\,(12)$&$7.1\pm0.9$&$2.0^{+2.1}_{-0.8}\,(-10)$&(4)\\[2pt]
CH$_3$OH-E&$1.2^{+0.8}_{-0.5}\,(13)$&$4.7\pm0.6$&$4.0^{+6.3}_{-2.1}\,(-10)$&(4)\\[2pt]
CH$_3$CHO&$3.6^{+1.0}_{-0.5}\,(12)$&$6.1\pm1.4$&$1.2^{+1.2}_{-0.4}\,(-10)$&(1)\\[2pt]
$t$-CH$_3$CH$_2$OH&$<1.2\,(13)$&$6.1\pm1.4$&$<5.0\,(-10)$&(1)\\[2pt]
CS&$>3.8\,(13)$&$6.1\pm1.4$&$>9.5\,(-11)$&(1)\\[2pt]
CCS&$1.3^{+0.2}_{-0.1}\,(13)$&$5.6\pm0.2$&$4.2^{+3.3}_{-1.5}\,(-10)$&(4)\\[2pt]
CC$^{34}$S&$7.5^{+2.7}_{-1.5}\,(11)$&$6.1\pm1.4$&$2.5^{+2.9}_{-1.0}\,(-11)$&(1)\\[2pt]
C$_3$S&$2.3^{+0.9}_{-0.4}\,(12)$&$6.1\pm1.4$&$7.6^{+9.1}_{-2.9}\,(-11)$&(1)\\[2pt]
\enddata
\tablecomments{Base-ten exponents are given in parentheses. Column density upper limits are $3\sigma$. Abundances are given with respect to H$_2$ column density $(3\pm1)\times10^{22}$~cm$^{-2}$.}
\tablenotetext{a}{Method used for column density calculation: (1) LTE with $T_{rot}=6.1\pm1.4$; (2) Rotational diagram; (3) Best-fit RADEX radiative transfer model (with density set to $n_{\rm H_2}=10^6$ cm$^{-3}$); (4) Rotational diagram including supplementary data from \citep{kon00}.}
\tablenotetext{b}{Total C$_4$H$_2$ column density calculated from ortho species assuming an equilibrium ortho-to-para ratio (3:1).}
\end{deluxetable}

\subsubsection{Rotational diagrams}
\label{sec:rotdiag}

Rotational diagrams \citep[see, \eg][]{cum86} were used for those species for which at least two transitions from significantly different upper-state energies were observed. The rotational diagrams containing at least three transitions each are shown in Figure \ref{fig:rotdiags}.  Where possible, in order to better constrain the temperatures and column densities, our data were supplemented with data from the SEST observations of  \citet{kon00}. Derived rotational excitation temperatures $T_{rot}$ are relatively low ($\lesssim10$~K), so it was necessary to account for the cosmic microwave background (CMB) radiation by including a factor $[1-(J_{\nu}(T_{bg})/J_{\nu}(T_{rot})]^{-1}$ in the column density calculations, where $J_{\nu}(T)$ is the Planck radiation law and $T_{bg}=2.73$~K is the CMB temperature. For HC$_3$N, C$_4$H, ortho-$c$-C$_3$H$_2$ and CCS, the peak opacities ($\tau_\nu$) are significant for one or more lines, so for these species the points on the rotational diagram were corrected using the factor

\be
\label{equ:cor}
c_\nu=\tau_\nu/(1-e^{-\tau_\nu}), 
\ee

where $\tau_\nu$ was derived from the opacity equation

\be
T_{\rm MB}=[J_{\nu}(T_{rot})-J_{\nu}(T_{bg})](1-e^{-\tau_\nu})
\label{equ:opacity}
\ee

The correction factors were determined iteratively, with initial values ($c_\nu^0$) calculated at temperatures of $T_{rot}^0({\rm HC_3N})=7.6$~K, $T_{rot}^0({\rm C_4H})=4.3$~K, $T_{rot}^0({\rm ortho}$-$c$-${\rm C_3H_2})=7.7$~K and $T_{rot}^0({\rm CCS})=5.8$~K, which were derived from the respective uncorrected rotational diagrams for these species. For each successive iteration, new correction factors ($c_\nu^n$) were calculated based on the temperatures derived from the previous iteration. Converged rotational temperatures and corresponding column densities are given in Table \ref{tab:colds}.

\subsubsection{LTE calculations}

For molecules for which rotational diagrams could not be plotted, column densities were calculated under the assumption of local thermodynamic equilibrium (LTE), using a gas temperature of $6.1\pm1.4$~K, which is the mean average ($\pm1\sigma$), of the excitation temperatures derived from the rotational diagrams. 

For molecules in this category with multiple line detections, the weighted-average column densities were calculated based on the individual line signal-to-noise ratios. Where molecular lines were searched for but not detected, column density upper limits were calculated using upper integrated line intensity limits of $3\Delta v\times({\rm RMS\ noise})$, for which $\Delta v=0.78$~\kms\ was used.  For C$_6$H$^-$, the upper limit of $4.79\times10^{10}$~cm$^{-2}$ was derived from the average of the three observed C$_6$H$^-$ spectra.

\subsubsection{Radiative transfer modelling}
\label{sec:radex}

\begin{figure*}
\centering
\includegraphics[width=0.75\columnwidth]{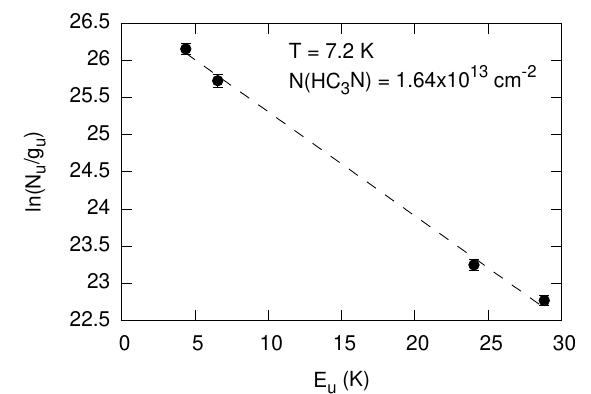}
\includegraphics[width=0.75\columnwidth]{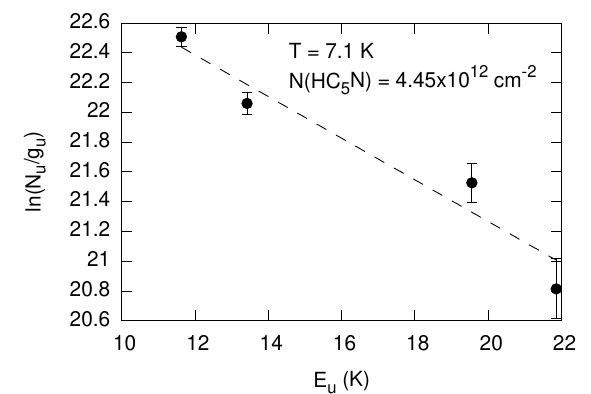}
\includegraphics[width=0.75\columnwidth]{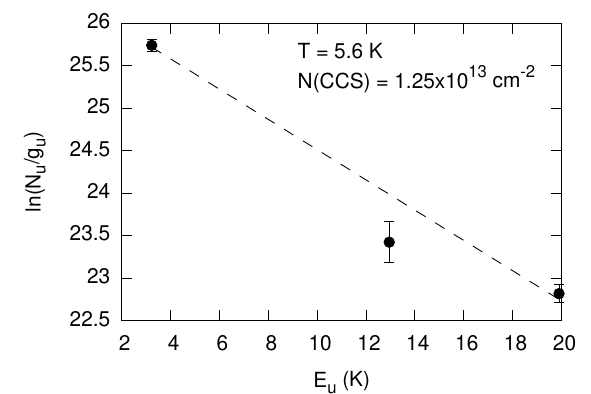}
\includegraphics[width=0.75\columnwidth]{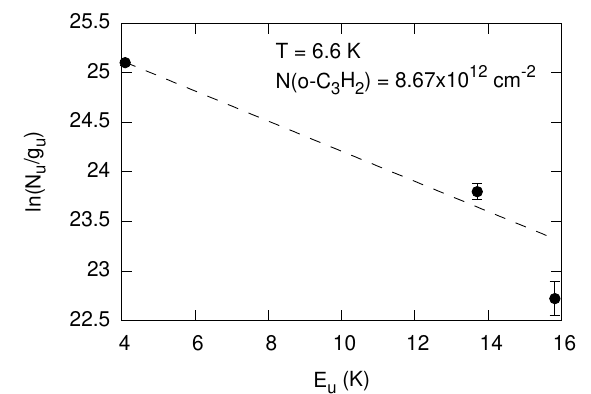}
\caption{Rotational diagrams for HC$_3$N, HC$_5$N, CCS and o-$c$-C$_3$H$_2$, including data from the present work and from \citet{kon00}.  Lines of best fit are shown (with corresponding LTE excitation temperatures $T$ and column densities $N$). Integrated line intensities have been opacity-corrected (see Section \ref{sec:rotdiag}).}
\label{fig:rotdiags}
\end{figure*}

For HC$_3$N and $c$-C$_3$H$_2$, the observed emission lines were subject to radiative transfer modelling using the RADEX code developed by \citet{van07}.  This routine employs a statistical equilibrium calculation for molecular excitation involving collisional and radiative processes, and accounts for optical depth effects using an escape probability method. Collisional and radiative (de-)excitation rates were taken from the Leiden Atomic and Molecular Database (LAMDA) \citep{sch05}\footnote{http://www.strw.leidenuniv.nl/$\sim$moldata}, which tabulates scaled versions of the original data published by \citet{gre78} for HC$_3$N and \citet{cha00} for $c$-C$_3$H$_2$. The RADEX model free parameters (number density of primary collision partner $n_{H_2}$, gas kinetic temperature $T$ and molecular column density $N$) were optimised using a least-squares algorithm to produce the best fit to the observed integrated line intensities (including the additional higher-frequency line data from \citealt{kon00}). 

The two nuclear-spin species of $c$-C$_3$H$_2$ (ortho and para) were considered separately. For ortho-$c$-C$_3$H$_2$, the following transitions were included in the fitting: $J_{K_aK_c}=2_{12}-1_{01}$, $3_{12}-2_{21}$ and $3_{21}-3_{12}$.  The best-fitting density was $1.0\times10^6$~cm$^{-3}$, but $T$ and $N$ were not well constrained by the data. Therefore, in addition to the LTE values, RADEX column densities for ortho- and para-$c$-C$_3$H$_2$ are given in Table \ref{tab:colds}, calculated using a fixed density of $n_{H_2}=1.0\times10^6$~cm$^{-3}$ and temperature $T=6.1\pm1.4$~K. Even at densities $\sim10^6$, the complexity of the $c$-C$_3$H$_2$ spectrum results in significant departures from LTE excitation, so the results from radiative transfer modelling are preferred over the LTE /rotational diagram results.  However, the calculated ortho-$c$-C$_3$H$_2$ energy level populations are highly dependant on the assumed temperature, which leads to substantial uncertainty in the column density.  On the other hand, the HC$_3$N rotational excitation is close to LTE, as shown by the near-linearity of the corrected rotational diagram in Figure \ref{fig:rotdiags}. As a result, the gas density is only constrained by a lower limit ($n_{H_2}\gtrsim5.5\times10^5$~cm$^{-3}$), derived using a RADEX fit to the $J=4-3$, $5-4$, $10-9$ and $11-10$ transitions. Above this value, the best-fitting HC$_3$N column density is relatively density-insensitive; the value for $N({\rm HC_3N})$ shown in Table \ref{tab:colds} was calculated for $n_{H_2}=1.0\times10^6$~cm$^{-3}$.  The best-fitting HC$_3$N excitation temperature was 7.1~K, which matches well the value of $7.2\pm0.2$~K derived from the corrected HC$_3$N rotational diagram. 

\section{Discussion}
\label{sec:dis}

\subsection{Molecular column densities and abundances}

Based on the dust emission, \citet{ten06} calculated an average H$_2$ column density of $N({\rm H_2})=(3\pm1)\times10^{22}$~cm$^{-2}$ in a $50''$ beam surrounding Cha-MMS1. Using this range, the observed molecular abundances relative to H$_2$ are given in Table \ref{tab:colds}.

Our column densities for CCS, $c$-C$_3$H$_2$, HC$_3$N and CH$_3$OH differ from those published by \citet{kon00}, who used a line-ratio method for the derivation of LTE excitation temperatures.  Their derived temperatures are significantly less than ours (for all species except A-methanol), leading to greater line opacities and column densities. For example, \citet{kon00} derived  $T_{ex}=4.7\pm0.2$~K, $N=4.7\times10^{13}$~cm$^{-2}$ for $c$-C$_3$H$_2$, and $T_{ex}=4.1\pm0.1$~K, $N=4.5\times10^{14}$~cm$^{-2}$ for HC$_3$N.  For their strongest observed lines of these species, core opacities of $\tau_\nu\approx3$ were derived, whereas we derive opacities less than 1 based on their data for the same transitions. Their HC$_3$N column density, in particular, seems unrealistically large compared to other carbon-chain rich interstellar clouds (even the TMC-1 cyanpolyyne peak); we derive values of \cold{2.0}{13} using RADEX modelling and \cold{1.6}{13} using a (corrected) rotational diagram.  Their use of approximate partition functions for ortho and para $c$-C$_3$H$_2$ is another source of discrepancy.

The mean gas number density within a ${\rm FWHM}=50''$ beam surrounding Cha-MMS1 was found by \citet{bel11} to be $9.8\times10^5$~cm$^{-3}$ (assuming a dust temperature of 12~K), which closely matches our derived H$_2$ number density of $n_{H_2}=1.0\times10^6$~cm$^{-3}$. If the dust temperature is closer to our mean derived value of 6.1~K, the density calculated by \citet{bel11} could be up to about a factor of 3 greater. This would still be consistent with our results however, given their smaller beam size, which probed more of the higher-density material nearer to the protostar.

\subsubsection{Polyynes and cyanopolyynes}
\label{sec:chains}

The abundances of carbon-chain-bearing species (including the polyynes and cyanopolyynes) in Cha-MMS1 are large, and similar to those found in other carbon-chain-rich interstellar clouds, prestellar cores and protostars such as TMC-1, L1527 \citep{sak08}, L1521E \citep{hir04}, L1512 and L1251A \citep{cor11}.

Previous laboratory and theoretical studies have shown that polyynes and cyanopolyynes may be rapidly synthesised in cold, dark interstellar clouds through gas-phase ion-molecule and neutral-neutral chemistry \citep[\eg][]{her89,smi04}. Large abundances of carbon-chain-bearing molecules are often taken as a sign of chemical youth in molecular clouds \citep[\eg][]{suz92,hir09,tak11}. Indeed, the large abundances of these species in TMC-1 can be explained as being due to the `early-time peak' that occurs when reactive carbon is still freely available in the gas (before it becomes almost completely locked up in CO). In environments denser than TMC-1, gas-phase species freeze out more rapidly onto dust due to increased collision rates, and this can become an important driver of large-molecule chemistry: Assuming equal sticking probabilities for all species involved in gas-grain collisions, lighter species (including atomic oxygen) travel at higher velocities and therefore freeze out more quickly than the heavier, slower-moving carbon chains. Because O is a primary destructive reactant for carbon-chain-bearing species, their abundances rise as oxygen freezes out.  This is known as the `freeze-out peak' \citep[\eg][]{bro90}, and may be a plausible explanation for the large carbon chain abundances observed in Cha-MMS1 and other dense cores. Evidence in favour of a freeze-out-peak chemistry in Cha-MMS1 is given by the CO depletion and large NH$_3$ abundance \citep{ten06}.

Alternatively, \citet{sak08,sak09} hypothesised that elevated carbon chain abundances in L1527 arise as a result of heating by the newly-formed protostar IRAS 04368+2557. Once the outer envelope reaches a temperature $\gtrsim30$~K, it is theorised that grain-surface methane begins to sublimate, which then reacts with gas-phase C$^+$ to form hydrocarbon ions. These subsequently engage in ion-molecule reactions and give rise to a so-called `warm carbon-chain chemistry' (WCCC). Chemical models show that this mechanism is capable of producing elevated abundances of unsaturated hydrocarbons including polyynes and cyanopolyynes in the warm regions surrounding low-mass protostars \citep{has08}. The detection of high-excitation-energy carbon chain emission lines in L1527 by \citet{sak08} is consistent with this theory. High-excitation lines have not yet been detected in Cha-MMS1, but we are presently undertaking new observations to search for them. The HC$_3$N $J=10-9$ map of Cha-MMS1 published by \citet{kon00} shows a north-south elongation in the peak emission contour, which covers both the location of the protostar and the more southerly peak in our observed HC$_3$N $J=4-3$ map. The $J=4-3$ emission originates from a rotational energy level ($E_u$) only 4.3~K above the ground state, whereas the $J=10-9$ emission originates from $E_u=24$~K and therefore provides evidence for a significant contribution to the HC$_3$N emission from warmer, more excited gas heated by the protostar. It is possible that carbon chain abundances may be enhanced as a result of WCCC in this region. However, \citet{kon00} did not detect any emission from the high-excitation ($E_u=52$~K) $J=15-14$ line of HC$_3$N, which indicates that any warm and dense region must be small compared with the $37''$ SEST beam.

According to the  physical model of protostar evolution by \citet{mas00}, during the FHSC stage soon before the protostar begins fusion, the temperature reaches above 30~K at radial distances from the core $R_{30}\lesssim20$~AU.  The protostar subsequently ignites and warms the surrounding envelope over time. After $2\times10^4$~yr, $R_{30}$ is predicted to increase to $\sim1000$~AU, and increases further during the mid-to-late stages of the Class 0 phase, but by a relatively smaller amount.  The Mopra beam HWHM of $\approx40''$ corresponds to 6000~AU at the distance of Cha-MMS1, so emission from the cool, extended outer envelope of the (collapsing) protostar dominates the observed spectra, and any emission from the compact methane-sublimation region (inside $R_{30}$) would be severely diluted regardless of the evolutionary state of the protostar, making the WCCC signature difficult to detect. In the case of Cha-MMS1 (with its very low luminosity), the warm region close to the protostar is probably too small to make a significant contribution to our observed emission. Our non-detection of the more highly-excited $^2\Pi_{1/2}\ J=27/2-25/2$ emission lines of C$_6$H (with $E_u=31$~K) is not surprising given the expected beam dilution. 

On the other hand, if Cha-MMS1 contains a VeLLO undergoing episodic accretion, the size of the warm (methane-sublimation) region is predicted to vary in accordance with the accretion rate \citep[see for example,][]{lee07}. Therefore, if Cha-MMS1 is presently at or near to a minimum in the accretion rate cycle, there may still be abundant carbon chains around the protostar, left over from a previous epoch of WCCC that occurred when the accretion rate and protostar luminosity were greater, and the size of $R_{30}$ correspondingly larger. It must be emphasised, however, that the `cold' carbon chain emission peaks (see Figures \ref{fig:hc3n_map} to \ref{fig:c3s_map}) are offset from the protostar centre, so WCCC is an unlikely explanation for the majority of the carbon-chain-rich material surrounding Cha-MMS1.  Smaller telescope beam sizes combined with the observation of high excitation-energy lines to probe warmer gases will be required for further analysis of the carbon chemistry inside the putative methane sublimation zone and warm inner envelope of Cha-MMS1.

\subsubsection{Anions}

The polyyne anions C$_4$H$^-$ and C$_6$H$^-$ were not detected. Three-sigma upper limits on their respective anion-to-neutral ratios are [C$_4$H$^-$]/[C$_4$H]~$<$~0.02\% and [C$_6$H$^-$]/[C$_6$H]~$<$~10\%, which include the errors in the observed abundances of the neutral species. The largest anion-to-neutral ratios observed in any astronomical source so far (for both species) have been in L1527 where [C$_4$H$^-$]/[C$_4$H]~=~0.01\% \citep{agu08} and [C$_6$H$^-$]/[C$_6$H]~=~9\% \citep{sak08c}, which are consistent with our results.

In chemical models for dense clouds \citep[\eg][]{mil07,har08}, the observed C$_6$H$^-$ anion-to-neutral ratios (on the order of a few percent) are reproduced with good accuracy. However, our non-detection of C$_4$H$^-$ provides a new example of the observed C$_4$H anion-to-neutral ratio being substantially less than predicted by theory \citep[see][]{her08}.

\subsubsection{Methanol}
\label{sec:methanol}

Methanol is found in the gas phase in dense molecular clouds, prestellar cores and in the vicinity of protostars, with column densities $\sim10^{13}-10^{15}$~cm$^{-2}$ \citep[\eg][]{fri88,mar05,buc06}.  However, most of the methanol in cold, dense protostellar envelopes is found in the form of ice \citep{boo08}. Its formation \emph{via} gas-phase chemistry is inefficient \citep{gep06}, and methanol is widely believed to be formed by successive hydrogenation of CO on cold dust grain surfaces.  Its appearance in the gas phase in cold regions has been hypothesised to be due to cosmic-ray-induced desorption, reactive desorption, or heating in grain-grain collisions, whereas inside protostellar envelopes, heating of the material to temperatures $\gtrsim100$~K results in complete evaporation of the ices \citep[\eg][]{rod03}.

The CH$_3$OH map in Figure \ref{fig:ch3oh_map} does not rule out the existence of methanol in warm gas in the vicinity of the protostar, but similar to the carbon-chain-bearing species, the fact that the emission peak is offset by $\sim10^4$~AU shows that protostellar heating probably does not explain the majority of the observed methanol. The lack of spatial correlation between the CH$_3$OH and carbon chain distributions matches the observations made by \citet{buc06}, who used chemical models to infer the physical and chemical histories of HC$_3$N-rich and CH$_3$OH-rich gas clumps. The observed methanol peak offset in Cha-MMS1 is consistent with the theory presented by \citet{mar00} and \citet{buc06}, that Alfv{\'e}n waves emitted by nearby collapsing protostars cause transient removal of ice mantles from the surrounding dust, resulting in the liberation of methanol into the gas phase at a distance from the protostellar core on the order of the MHD-wave damping length ($\sim10^4$~AU). \citet{mar00} used this theory to successfully explain the origin of the abundance gradients of various species observed along the ridge of TMC-1. The young age of Cha-MMS1 implies that protostellar collapse is currently in progress, resulting in the production of Alfv{\'e}n waves that could be responsible for desorption of the observed methanol.

From the ratios of our observed CH$_3$OH lines with those of \citet{kon00}, differing temperatures were calculated for the A and E methanol nuclear spin species of $7.1\pm0.9$~K and $4.7\pm0.6$~K, respectively. The column densities of A and E methanol (given in Table \ref{tab:colds}) also differ, but within the error bars. The ratio of the sums over the statistical weights of the quantum states of A and E methanol approaches unity, so in the limit of high temperature (in thermodynamic and chemical equilibrium), the abundances of A and E methanol should be approximately equal. Convergence of the A and E abundances is calculated to occur for temperatures above about 30~K \citep[see][]{wir09,wir11}. At lower temperatures, A-methanol is the more abundant species because its ground state is 7.9~K lower in energy than that of E-methanol.  Given that both spin species are likely to have the same spatial distribution in the ISM, their relative column densities in Cha-MMS1 may be more accurately compared by assuming a common excitation temperature.  In LTE at 6.1~K, the column densities of A and E methanol calculated from our observations are $(7.2\pm0.7)\times10^{12}$~cm$^{-2}$ and $(6.6\pm0.7)\times10^{12}$~cm$^{-2}$.  Using only the higher-frequency lines observed by \citet{kon00}, we calculate the respective column densities to be $(7.5\pm0.7)\times10^{12}$~cm$^{-2}$ and $(7.9\pm0.8)\times10^{12}$~cm$^{-2}$.  These results are consistent with equal abundances of A and E methanol, implying formation at close to statistical equilibrium.

\subsubsection{Deuterium fractionation}

In cold interstellar clouds, molecular ions become enriched in deuterium as a result of exothermic gas-phase fractionation reactions involving HD \citep{mil89}.

\citet{bel06} found a deuterated N$_2$H$^+$ fraction in Cha-MMS1 of [N$_2$D$^+$]/[N$_2$H$^+$]~=~11\%, which is typical for prestellar cores and young protostars observed in the nearby Galaxy. The cyanoacetylene deuteration fraction [DC$_3$N]/[HC$_3$N]~=~$3.6\pm1.5$\% is within the range of values previously observed in TMC-1 ($1-6$\%; \citealt{tur01,sai02}), and is similar to the carbon-chain-rich prestellar core L1544 (6\%; \citealt{how94}), and the protostar L1527 (3\%; \citealt{sak09b}). {These results are consistent with chemical models for cold interstellar clouds \citep[\eg][]{rob00,tur01,pag09}.}

\subsubsection{Cyclopropenylidene}
\label{sec:c3hd}

A large $c$-C$_3$HD column density of \cold{4.5^{+1.3}_{-0.7}}{12} is found in Cha-MMS1, which is about 5 times greater than in L1527 and an order of magnitude greater than in TMC-1 \citep{sak09b,tur01}. Accounting for the rather large uncertainty in the $c$-C$_3$H$_2$ column density (see Section \ref{sec:radex}), we calculate that the cyclopropenylidene deuteration fraction [$c$-C$_3$HD]/[$c$-C$_3$H$_2$] is between 5\% and 34\%. This is not significantly different from the range of values ($5-15$\%) reported in a sample of twelve dark clouds by \citet{bel88}, and the value of 7\% reported in L1527 by \citet{sak09b}, but is consistent with previous results which show the $c$-C$_3$H$_2$ deuteration fraction is larger than that of HC$_3$N \citep[\eg][]{tur01,sak09b}. Chemical models predict comparable levels of deuteration for $c$-C$_3$HD and HC$_3$N; deuterium enrichment in both these species is considered to originate predominantly from the CH$_2$D$^+$ ion \citep[\eg][]{rob00,tur01}. Deuteration may also occur by deuteron transfer as a result of collisions with H$_2$D$^+$, followed by dehydrogenation upon recombination. Discrepancy between the deuteration fractions of $c$-C$_3$HD and HC$_3$N is indicative of an incomplete understanding of the deuterium chemistry of these two species. The fact that $c$-C$_3$H$_2$ can become doubly deuterated may partly explain why $c$-C$_3$HD is so abundant: if sufficient $c$-C$_3$D$_2$ forms, it would act as a buffer for $c$-C$_3$HD in subsequent proton-deuteron exchange processes. It may thus be of interest to search for $c$-C$_3$D$_2$ in Cha-MMS1 and similar environments. However, we are not aware of any published laboratory studies on the rotational spectrum of $c$-C$_3$D$_2$; such a study would be warranted. 

The $c$-C$_3$H$_2$ molecule exists in two nuclear spin configurations: ortho and para, and it has been suggested that the ortho-to-para ratio (OPR) may be used as a probe of interstellar cloud ages \citep{mor06}. However, our present data are insufficient to usefully constrain the OPR. Additional observations of $c$-C$_3$H$_2$ and $c$-C$_3$HD lines from a range of energy energy levels and with a range of optical depths will be required in order to better constrain the column densities of these species and derive more reliable values for the $c$-C$_3$H$_2$ OPR and deuteration fraction.

\subsubsection{Sulphuretted species}

We observed large abundances of the sulphur-bearing carbon chains C$_n$S ($n=1-3$), on the same order of magnitude as found in various parts of the Taurus molecular cloud complex \citep[\eg][]{hir04}. Our [CCS]/[C$_3$S] column density ratio ($5.7\pm2.1$) is greater than the values observed by \citeauthor{hir04}, of 2.1, 2.3, 2.0 in the dense cores L1495B, L1521B and L1521E, respectively, and is more similar to the value of 5.1 found in TMC-1. The relatively large value in Cha-MMS1 is primarily attributable to a relatively low C$_3$S column density, which is more consistent with the lower C$_3$S column densities observed in the carbon-chain-rich cloud cores L1512 and L1251A by \citet{cor11}.  

Sulphur is believed to be incorporated into gas-phase hydrocarbon molecules in the ISM mainly by ion-neutral chemistry involving S$^+$ \citep{mil90}. Recombination of HC$_n$S$^+$ species then produces sulphuretted carbon chains. The main source of HC$_3$S$^+$ is from the reaction S$^+$ + $c$-C$_3$H$_2$ $\longrightarrow$ HC$_3$S$^+$ + H, followed by recombination of HC$_3$S$^+$ to produce C$_3$S. Thus, the relatively low C$_3$S abundance in Cha-MMS1 could be explained by a relative underabundance of $c$-C$_3$H$_2$ compared with L1495B, L1521B and L1521E. An alternative method for the synthesis of C$_3$S is by the neutral-neutral reaction C$_2$H + CS $\longrightarrow$ C$_3$S + H \citep{smi04}. However, this reaction is relatively slow (two orders of magnitude slower than the ion-neutral reaction), and is therefore expected to be less important in the production of C$_3$S.

The [N$_2$H$^+$]/[CCS] ratio may be used as a measure of chemical age in interstellar clouds on the basis that carbon chains are most abundant at early times, whereas N$_2$H$^+$ becomes more abundant later on as CO freezes out \citep[\eg][]{tak11}. Using the N$_2$H$^+$ abundance from \citet{bel06}, we derive [N$_2$H$^+$]/[CCS]~=~$0.5\pm0.3$ in Cha-MMS1, which is similar to the value of $0.52\pm0.07$ found in the young, carbon-chain-rich VeLLO L1521F \citep{tak11}, indicating a comparable chemical age, but less than the value $1.18\pm0.21$ in the more evolved VeLLO IRAM~04191.

Only a lower limit could be placed on the CS column density because the optical depth of the $J=1-0$ lines approaches infinity for the adopted temperature range of $6.1\pm1.4$~K.

\subsubsection{Other molecules}
\label{sec:other}

Acetaldehyde (CH$_3$CHO) was detected in Cha-MMS1 with a column density of \cold{3.6}{12}. It is an important molecule in terrestrial organic chemistry, and has been detected in star-forming regions, protostars and translucent/dense molecular clouds. A range of gas and ice-phase reaction pathways have been proposed to explain its abundance in space \citep[see][and references therein]{cha04,ben05}. However, the size of the Mopra beam does not place any useful constraint on the origin of the emission relative to the protostar, and therefore, apart from highlighting the diversity of the organic chemistry of this region, the observed CH$_3$CHO abundance of $\approx10^{-10}$ does not significantly constrain the physics or chemistry of the molecule or the protostar.  Our observations are, however, consistent with equal abundances of gas-phase A- and E- CH$_3$CHO, indicating statistical equilibrium.

Ethanol (CH$_3$CH$_2$OH) has previously been observed with relatively large abundances of $\sim10^{-8}$ in `hot core' regions surrounding high-mass protostars \citep[\eg][]{mil95}, but is not present in quiescent (non-star-forming) molecular clouds. There have been two reported detections to-date in low-mass protostars: an abundance of $\sim10^{-8}$ was found in the hot corino IRAS 16293-2422A by \citet{hua05} and $\sim10^{-7}$ in the L1157 outflow by \citet{arc08}. We derive a (beam-averaged) upper limit on the ethanol abundance of $5\times10^{-10}$ in Cha-MMS1.

We also searched for the protonated carbon dioxide ion HCO$_2^+$ and derive an upper abundance limit of $4.7\times10^{-11}$.  This molecule was detected by \citet{sak08b} in the vicinity of the low-mass protostar L1527, with a beam-averaged abundance of $1.3\times10^{-12}$. HCO$_2^+$ is believed to be formed predominantly by protonation of CO$_2$, and the gas-phase CO$_2$ abundance can be estimated using Equation 5 of \citet{sak08b}. Using a CO abundance of $5\times10^{-5}$ (calculated using the C$^{18}$O column density of $2.5\times10^{15}$~cm$^{-2}$ reported by \citealt{kon00} and an intrinsic $^{16}$O/$^{18}$O ratio of 500), we derive an upper limit of $n({\rm CO}_2)<4\times10^{-6}n_{\rm H_2}$, averaged over the Mopra beam.  Spitzer observations \citep{fur08} show a solid CO$_2$ abundance of $5\times10^{-6}$ towards L1527 and a similar ice abundance may be expected in the cool, outer envelope of Cha-MMS1. The relatively low gas-phase CO$_2$ in Cha-MMS1 implied by our observations, combined with the upper limit on the ethanol abundance, shows that the energetic processing of matter that would cause the sublimation of dust grain mantles in the protostellar envelope, due to heating by the protostar or an outflow shock, for example, must be lacking in Cha-MMS1, or at least confined to a relatively small area compared with the $\approx90''$ Mopra beam.

\subsection{Comparison with other sources}

The molecular abundances in Cha-MMS1 are indicative of an active (non-equilibrium) carbon chemistry, resulting in the synthesis of large quantities of unsaturated carbon-chain-bearing species such as the polyynes and cyanopolyynes. The carbon-chain abundances are similar to those in young, quiescent cores such as TMC-1 and the `carbon-chain-producing regions' observed by \citet{suz92} and \citet{hir04,hir09}. However, as pointed out by \citet{ten06}, the CO depletion and large NH$_3$ abundance in Cha-MMS1 are hallmarks of a later stage of chemical evolution, associated with the freezing-out of gas-phase species onto the dust, as has been observed in prestellar cores and the outer envelopes of other young, low-to-intermediate-mass protostars \citep{cas99,jor04,cra05,alo10}. 

Only a few detections of long carbon-chain-bearing species in low-mass protostellar envelopes have been reported to-date. The smaller species HC$_3$N, C$_3$H$_2$ and C$_4$H in Cha-MMS1 are a factor of a few to an order of magnitude more abundant than those surrounding the VeLLO L1521F and the prestellar core L1544 \citep[see][]{suz92,gup09,tak11}, both of which are considered to be relatively carbon-chain-rich sources. Due to their evolved state, the appearance of carbon-chain-bearing molecules in these sources is likely to be the result of `freeze-out-peak' chemistry (see Section \ref{sec:chains}).

Carbon chains have also been hypothesised to arise in large abundances in protostellar envelopes as a result of warm carbon-chain chemistry (WCCC) \citep{sak09}. Of all the dense molecular clouds in the literature for which multiple complex molecules have been observed, the abundances in Cha-MMS1 match most closely with those in the well-studied Class~0/I protostar L1527 in the Taurus molecular cloud complex \citep{sak08,sak09,sak09b}, where WCCC has been theorised to be occurring. Indeed, the abundances of all observed species in Cha-MMS1 are within an order of magnitude of those observed in L1527. 

The gas densities in the envelopes of Cha-MMS1 and L1527 are quite similar ($\sim10^6$~cm$^{-3}$), so the similarities in their chemistries may be related to similarities in the physical conditions. It is tempting, therefore, to hypothesise that the large abundances of organic molecules around Cha-MMS1 also arise as a result of WCCC. However, Cha-MMS1 is likely to be at a significantly earlier stage of evolution than L1527, and has an estimated luminosity $\sim1/100$ as large \citep{jor02,bel11}. The low IR flux and lack of outflow from the protostar \citep{bel06,hir07} indicate that Cha-MMS1 is very close to the beginning of the Class~0 phase, whereas L1527 is more evolved, with bright scattered IR emission and clear outflow cavities, consistent with an edge-on Class~I protostar obscured by its own disk \citep{tob08}. Assuming a uniform accretion rate, \citet{oha97} calculated the age of the L1527 protostar to be $\sim10^5$~yr, which would place it at the end of the Class 0 phase \citep[based the Class 0 lifetime derived by][]{eno09}.  Chemical timescales in dense interstellar clouds (in particular, molecular freeze-out timescales; see Equation 4 of \citealt{buc06}) are significantly less than the estimated $\sim10^5$~yr age difference of these two protostars. Therefore, the chemical similarity between Cha-MMS1 and L1527 may be coincidental. Notwithstanding the possibility of episodic outbursts, the low luminosity of Cha-MMS1 implies a very small hot-core region (probably $\lesssim10$~AU in size), that would not be detectable with single-dish microwave telescopes. Instead, the large carbon chain abundances observed in the envelope of Cha-MMS1 may be present because this young protostar is forming from a very chemically-rich parent molecular cloud.

\citet{cra05} derived CO depletion factors $\delta_{\rm CO}\sim10$ in L1521F and L1544. Using the CO abundance from Section \ref{sec:other}, we calculate $\delta_{\rm CO}\sim2$ in Cha-MMS1 (assuming an undepleted CO abundance of $9.5\times10^{-5}$ after \citealt{cra05}). The size of the warm region around the Cha-MMS1 protostar within which CO ice is desorbed is likely to be negligible compared with the telescope beam, so low CO depletion may be a sign that the freezing-out of gas-phase species onto the dust is less far advanced. This would be consistent with a relatively short prestellar phase for Cha-MMS1. Alternatively, the low CO depletion could be a consequence of grain mantle destruction; significant turbulent motion of the gas around Cha-MMS1 may be inferred from the relatively broad molecular linewidths ($\approx0.78$~\kms), which are more than an order of magnitude greater than the thermal limit in most cases, and are significantly broader than observed in L1521F ($<0.3$~\kms; \citealt{gup09}). This may be a signature of energetic cloud motions that cause collisions between dust grains, resulting in the liberation of mantle species into the gas phase.  Possible physical processes responsible for mantle destruction include inter-clump collisions or Alfv{\'e}n waves emitted by the protostar. As discussed in Section \ref{sec:methanol}, the liberation of mantle species may explain the origin of the observed methanol. According to \citet{mar00}, grain mantle desorption also gives rise to large gas-phase abundances of polyynes and cyanopolyynes.

To help ascertain the origin of complex molecules in the vicinity of Cha-MMS1, direct observations of the core region at higher spatial resolution are required, which will also help to establish the precise physical nature of the source and identify molecular signatures of the earliest stages of star formation. For example, observations of the gas-phase products of grain-surface chemistry (in particular, alcohols, saturated organic molecules and deuterated species) will probe the presence of a possible hot corino \citep[\eg][]{caz03}. High-resolution maps will permit the search for molecular abundance gradients due to Alfv{\'e}n waves predicted by \citet{mar00}. The identification of concentrated abundances of carbon chains around the protostar would provide evidence for warm carbon chain chemistry \citep{sak10b}. \citet{lee07} theorised that by mapping the N$_2$H$^+$ and CO abundances at high angular resolution, and using infrared CO$_2$ ice observations in the protostellar envelope, it should be possible to observe the chemical effects of the episodic accretion process hypothesised to be present in VeLLOs. \citet{omu07} calculated that high-excitation submillimeter H$_2$O emission lines may be used as a tracer for the presence of the infall shock-front that is predicted to be a characteristic property of first hydrostatic cores. A compact, low-velocity outflow is predicted to be another characteristic of the FHSC phase \citep{mac08,tom10}. It may thus be possible to determine whether Cha-MMS1 is a VeLLO or an FHSC using high angular-resolution ($\sim0.1-1''$) molecular line observations.

\section{Conclusion}
\label{sec:con}

The envelope of the low-luminosity protostar Cha-MMS1 exhibits an unusually rich chemical diversity. It contains very large abundances of organic molecules, indicative of active gas-phase and grain-surface chemical processes near to the protostar. The abundances of long carbon chains (including C$_6$H and HC$_7$N) are particularly large, and similar to those found in the most carbon-chain-rich regions of the Galaxy. These observations show that complex organic molecules and carbon-chain-bearing species can be abundant in star-forming gas, and may therefore be of relevance to studies of the chemical evolution of protoplanetary disks. The molecular abundances in Cha-MMS1 are similar to those in the envelope of the older, more luminous Class~0/I protostar L1527, and are greater than have been reported so far for any other low-luminosity, low-mass Class~0 protostellar envelope (including the VeLLOs L1521F, L1014 and IRAM~04191). This may be attributable to a combination of factors including (1) a young chemical age for the Cha-MMS1 natal interstellar cloud, (2) freezing out of atomic oxygen onto the dust, and (3) the liberation of grain mantle species into the gas by various mechanisms such as clump-clump collisions, Alfv{\'e}n waves or warming by the protostar.

The observations presented here are of lines from relatively low excitation levels which trace predominantly cool material; the chemistry of the warmer regions closer to the protostar may be examined by the observation of higher-excitation lines at higher spatial resolution. This will permit further analysis of the relationship between protostar and envelope chemistry. Establishing the molecular inventory of the envelope of Cha-MMS1 and other young protostars will be of fundamental importance towards the goal of understanding the initial chemical reagents available for the formation of planetary systems.

\acknowledgments

We gratefully acknowledge the assistance of Balt Indermuehle at Mopra for providing support during on-site and remote observations. Thanks to Arnaud Belloche for comments on the manuscript. This work was supported by the Goddard Center for Astrobiology and NASA's Origins of Solar Systems and Exobiology programs.

{\it Facilities:} \facility{Mopra}

\bibliographystyle{aa}
\bibliography{refs}

\end{document}